\documentclass[aps,pra,preprintnumbers,showpacs,tightenlines]{revtex4}
\usepackage{amssymb}
\usepackage{amsmath}
\usepackage{graphicx}
\usepackage{epsfig}
\usepackage{subfigure}
\usepackage{amsfonts}
\usepackage{CJK}

\begin{document}

\title{Circuit QED: Generation of two-transmon-qutrit entangled states via resonant interaction}

\author{Xi-Mei Ye $^{1}$, Zhen-Fei Zheng $^{3}$, Dao-Ming Lu $^{1}$, and Chui-Ping Yang $^{2}$}
\email{yangcp@hznu.edu.cn}
\address{$^1$College of Mechanic and Electronic Engineering, Wuyi  University, Wuyishan, Fujiang 354300, China}
\address{$^2$KDepartment of Physics, Hangzhou Normal University, Hangzhou 310036, China}
\address{$^3$Department of Physics, University of Science and Technology of China, Heifei 230026, China}

\date{\today}

\begin{abstract}
We present a way to create entangled states of two superconducting transmon qutrits based on circuit QED.
Here, a qutrit refers to a three-level quantum system. Since only resonant interaction is employed, the entanglement creation can be completed within a short time. The degree of entanglement for the prepared entangled state can be controlled by varying the weight factors of the initial state of one qutrit, which allows the prepared entangled state to change from a partially entangled state to a maximally entangled state. Because a single cavity is used, only resonant interaction is employed, and none of identical qutrit-cavity coupling constant, measurement, and auxiliary qutrit is needed, this proposal is easy to implement in experiments. The proposal is quite general and can be applied to prepare a two-qutrit partially or maximally entangled state with two natural or artificial atoms of a ladder-type level structure, coupled to an optical or microwave cavity.
\end{abstract}
\maketitle
\date{\today }

\begin{center}
\textbf{I. INTRODUCTION AND MOTIVATION}
\end{center}

Entanglement is one of the most striking aspects of quantum mechanics. A
system is entangled if its quantum state cannot be described as a direct
product of states of its subsystems. Entanglement has many applications in
quantum commuication and quantum information processing (QIP) [1-5]. In the
last decade, quantum entanglement engineering has been widely studied with
various physical systems. For instance, entanglement creation has been
theoretically proposed with superconducting (SC) qubits coupled via
Josephson junctions [6] or capacitors [7-9]. The focus of this work is on
entanglement creation with a circuit-QED system, which has been considered
as one of the leading candidates for QIP [10-17]. For the past years, based
on circuit QED, much process has been made on entanglement generation with
SC qubits. For instance, using SC qubits coupled to a single cavity or
multiple resonators (hereafter, the terms cavity and resonator are used
interchangeably), a number of theoretical proposals have been presented for
realizing entangled states of SC qubits [10-12,18-23]. Moreover, various
two-qubit or three-qubit entangled states have been experimentally
demonstrated with superconducting qubits based on circuit QED [24-29].
Recently, using SC qubits coupled to a single resonator, a
Greenberger-Horne-Zeilinger entangled state of ten SC qubits has been
demonstrated in experiments [30]. With solid-state platforms, this is the
largest number of entangled qubits reported so far, after the previous
experimental demonstration of entanglement with five superconducting qubits
via capacitance coupling was reported [31].

A significant experimental challenge for entanglement engineering is how to
create quantum entanglement with high-dimensional quantum systems. Compared
with a qubit (two-level system), a \textit{qutrit} (three-level system) has
a larger Hilbert space and thus can be used to encode more information in
QIP and communication. As is well known, high-dimensional (HD)\ quantum
entanglement is not only of great interest in providing an additional way
for the fundamental tests of quantum nonlocality, but also central for
HD-based quantum computation, quantum communication, quantum error
correction, and quantum simulation. Over the past years, base on cavity or
circuit QED, a large number of theoretical methods have been presented for
creating two-qubit or multi-qubit entangled states with various physical
systems (e.g., atoms, nitrogen-vacancy centers, quantum dots, and SC
devices) [10-12,18-23,32-39]. However, after a deep search of the
literature, we found that based on cavity QED or circuit QED, how to create
entangled states of qutrits or higher-dimensional quantum systems with
natural or artificial atoms has been rarely investigated.

Motivated by the above, in this work we will consider a physical system,
which consists of two superconducting transmon qutrits coupled to a single
microwave cavity or resonator. Note that transmon devices have been
considered as one of the best SC information carriers because of their
relatively longer coherence times compared with other type of SC devices
(e.g., flux, phase, and charge qubits/qutrits, etc.). In the following, we
will propose an approach to prepare the two transmon qutrits in an\
entangled state (either a partially-entangled state or a maximally-entangled
state). To the best of our knowledge, our proposal is the first one to show
how to create entangled states of qutrits with natural or artificial atoms
based on cavity or circuit QED.

As shown below, this proposal has the following advantages: (i) Because only
qutrit-cavity resonant interaction and qutrit-pulse resonant interaction are
employed, the entanglement creation can be completed within a short time;
(ii) There is no requirement on the identical qutrit-cavity coupling
constants; thus either non-uniformity in the qutrit device parameters
(resulting in nonidentical qutrit level spacings) or non-exact placement of
qutrits in the cavity is allowed by this proposal; (iii) Since only a single
microwave cavity is used and no auxiliary qutrit is needed, the setup is
simple; (iv) There is no need of measurement on the qutrit states or the
cavity state, thus the entangled states can be created deterministically;
and (v) The degree of entanglement for the prepared entangled state is
adjustable by varying the weight factors of the initial state of one qutrit,
thus this proposal can be used to create the two-qutrit entangled state
varying from a partially entangled state to a maximally entangled state.
This proposal is quite general, which can be extended to create a two-qutrit
entangled state with a wide range of physical systems, such as two natural
atoms (with a ladder-type three-level structure) coupled to an optical
cavity, or two artificial atoms of a ladder-type three-level structure
(e.g., superconducting phase qutrits, Xmon qutrits, nitrogen-vacancy
centers, quantum dots, etc.) coupled to a microwave cavity or resonator.

This paper is organized as follows. In Sec. II, we review the basic theory of
a transmon qutrit resonantly interacting with a single cavity or a classical
microwave pulse. In Sec. III, we explicitly show how to prepare two transmon
qutrits in an entangled state based on circuit QED. In Sec. IV, we give a
brief discussion on the experimental feasibility. A concluding summary is
given in Sec. V.

\begin{center}
\textbf{II. BASIC THEORY}
\end{center}

A transmon qutrit has a ladder-type three-level structure [40]. In this
work, we consider the three lowest levels of a transmon qutrit, which are
denoted as $\left\vert 0\right\rangle ,$ $\left\vert 1\right\rangle $ and $%
\left\vert 2\right\rangle $ (Fig.1). For a transmon qutrit, the transition
between the two levels $\left\vert 0\right\rangle $ and $\left\vert
2\right\rangle $ is much weaker compared to the transition between the two
levels $\left\vert 0\right\rangle $ and $\left\vert 1\right\rangle $ and the
transition between the two levels $\left\vert 1\right\rangle $ and $%
\left\vert 2\right\rangle $ [40]. As shown in next section, the entanglement
generation requires two types of resonant interaction. Namely, the resonant
interaction between a cavity and the $\left\vert 0\right\rangle
\leftrightarrow \left\vert 1\right\rangle $ transition of the qutrit; and
the resonant interaction between a classical pulse and the $\left\vert
1\right\rangle \leftrightarrow \left\vert 2\right\rangle $ transition of the
qutrit. To make our presentation given in next section clear, in the
following we will give a brief review on the state evolutions for these two
kinds of interactions.

\begin{figure}[tbp]
\begin{center}
\includegraphics[bb=170 438 290 569, width=3.5 cm, clip]{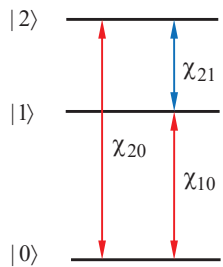} \vspace*{%
-0.08in}
\end{center}
\caption{Level diagram of a transmon qutrit with the ladder-type three
lowest levels. For a transmon qutrit, the level spacing between the upper
two levels is smaller than that between the two lowest levels. In addition,
for a transtion qutrit, the transition between the two levels $\left\vert
0\right\rangle $ and $\left\vert 2\right\rangle $ is much weaker compared to
the transition between the upper two levels and the transition between the
two lowest levels [34]. Namely, $\protect\chi _{20}\ll \protect\chi _{10},%
\protect\chi _{21}$. Here, $\protect\chi _{ij}$ is the transition matrix
element for the two levels $\left\vert j\right\rangle $ and $\left\vert
i\right\rangle $ ($ij=20,10,21$).}
\label{fig:1}
\end{figure}

\begin{center}
\textbf{A. Qutrit coupled to a single cavity}
\end{center}

Consider a transmon qutrit coupled to a single-mode cavity. Assume that the
cavity is resonantly coupled to the transition between the two\ lowest
levels $\left\vert 0\right\rangle $ and $\left\vert 1\right\rangle $ while
highly detunned (decoupled) from the transitions between other levels of the
qutrit. Under this consideration, the interaction Hamiltonian in the
interaction picture, after making the rotating-wave approximation, can be
written as $H_{\mathrm{I}_{1}}=\hbar ga^{\dagger }\left\vert 0\right\rangle
\left\langle 1\right\vert +$ h.c.$,$ where $a^{\dagger }$ ($a$) is the
creation (annihilation) operator of the cavity and $g$ is the coupling
constant between the cavity and the $\left\vert 0\right\rangle
\leftrightarrow \left\vert 1\right\rangle $ transition of the qutrit.

Under the Hamiltonian $H_{\mathrm{I}_{1}}$, one can obtain the following
state evolution:
\begin{eqnarray}
\left\vert 0\right\rangle \left\vert 0\right\rangle _{c} &\rightarrow
&\left\vert 0\right\rangle \left\vert 0\right\rangle _{c}  \notag \\
\left\vert 1\right\rangle \left\vert 0\right\rangle _{c} &\rightarrow
&-i\sin gt\left\vert 0\right\rangle \left\vert 1\right\rangle _{c}+\cos
gt\left\vert 1\right\rangle \left\vert 0\right\rangle _{c},  \notag \\
\left\vert 0\right\rangle \left\vert 1\right\rangle _{c} &\rightarrow &\cos
gt\left\vert 0\right\rangle \left\vert 1\right\rangle _{c}-i\sin
gt\left\vert 1\right\rangle \left\vert 0\right\rangle _{c},
\end{eqnarray}%
where $\left\vert 0\right\rangle _{c}$ is the vacuum state of the cavity
while $\left\vert 1\right\rangle _{c}$ is the single-photon state of the
cavity.

As shown in next section, the entanglement preparation requires that a
cavity interacts with each of two qutrits $1$ and $2$. In reality, it is a
challenge to have the qutrit-cavity coupling constants to be identical for
both of the qutrits. Thus, we replace the coupling constant $g$ with $g_{1}$
for qutrit $1$ while $g_{2}$ for qutrit $2.$\textbf{\ }

\begin{center}
\textbf{B. Qutrit driven by a classical pulse}
\end{center}

Let us consider a transmon qutrit driven by a microwave pulse. If the pulse
is resonantly coupled to the $\left\vert 0\right\rangle \leftrightarrow
\left\vert 1\right\rangle $ transition but highly detunned (decoupled) from
the transitions between other levels of the qutrit. Then, the interaction
Hamiltonian in the interaction picture is given by $H_{\mathrm{I}_{2}}=\hbar
\left( \mathrm{\Omega }_{10}e^{i\phi }\left\vert 0\right\rangle \left\langle
1\right\vert +\text{h.c.}\right) ,$ where $\phi $ is the initial phase of
the pulse while $\mathrm{\Omega }_{10}$ is the Rabi frequency of the pulse.
From the Hamiltonian $H_{\mathrm{I}_{2}}$, it is easy to show that a pulse
of duration $t$ results in the following state evolution:
\begin{eqnarray}
\left\vert 0\right\rangle  &\rightarrow &\cos \mathrm{\Omega }%
_{10}t\left\vert 0\right\rangle -ie^{-i\phi }\sin \mathrm{\Omega }%
_{10}t\left\vert 1\right\rangle ,  \notag \\
\left\vert 1\right\rangle  &\rightarrow &-ie^{i\phi }\sin \mathrm{\Omega }%
_{10}t\left\vert 0\right\rangle +\cos \mathrm{\Omega }_{10}t\left\vert
1\right\rangle .
\end{eqnarray}%
Similarly, when the pulse is resonantly coupled to the $\left\vert
1\right\rangle \leftrightarrow \left\vert 2\right\rangle $ transition but
highly detunned (decoupled) from the transitions between other levels of the
qutrit, the interaction Hamiltonian in the interaction picture is given by $%
H_{\mathrm{I}_{3}}=\hbar \left( \mathrm{\Omega }_{21}e^{i\phi }\left\vert
1\right\rangle \left\langle 2\right\vert +\text{h.c.}\right) ,$ where $%
\mathrm{\Omega }_{21}$ is the Rabi frequency of the pulse. Based on this
Hamiltonian, we can obtain the following state rotation:

\begin{eqnarray}
\left\vert 1\right\rangle  &\rightarrow &\cos \mathrm{\Omega }%
_{21}t\left\vert 1\right\rangle -ie^{-i\phi }\sin \mathrm{\Omega }%
_{21}t\left\vert 2\right\rangle ,  \notag \\
\left\vert 2\right\rangle  &\rightarrow &-ie^{i\phi }\sin \mathrm{\Omega }%
_{21}t\left\vert 1\right\rangle +\cos \mathrm{\Omega }_{21}t\left\vert
2\right\rangle .
\end{eqnarray}

Finally, it should be mentioned that when the cavity is highly detunned
(decoupled) from the $\left\vert 1\right\rangle \leftrightarrow \left\vert
2\right\rangle $ and $\left\vert 0\right\rangle \leftrightarrow \left\vert
2\right\rangle $ transitions of the qutrit, there is neither coupling
between the cavity and the qutrit nor transition between the levels $%
\left\vert 2\right\rangle $ and $\left\vert 1\right\rangle $ or $\left\vert
0\right\rangle $ induced by the cavity. As a result, the states $\left\vert
2\right\rangle \left\vert 0\right\rangle _{c},$ and $\left\vert
2\right\rangle \left\vert 1\right\rangle _{c}$ remain unchanged during the
cavity on resonance with the $\left\vert 0\right\rangle \leftrightarrow
\left\vert 1\right\rangle $ transition of the qutrit. For a similar reason,
the state $\left\vert 2\right\rangle $ ($\left\vert 0\right\rangle $) of the
qutrit is unaffected during applying the pulse resonant with the $\left\vert
0\right\rangle \leftrightarrow \left\vert 1\right\rangle $ ($\left\vert
1\right\rangle \leftrightarrow \left\vert 2\right\rangle $) transition of
the qutrit. These results, togeter with the results given in Eqs. (1-3),
will be employed for the entanglement generation, as shown in the next
section.

\begin{center}
\textbf{III. GENERATION OF TWO-TRANSMON-QUTRIT ENTANGLED STATES}
\end{center}

\begin{figure}[tbp]
\begin{center}
\includegraphics[bb=94 205 386 772, width=8.5 cm, clip]{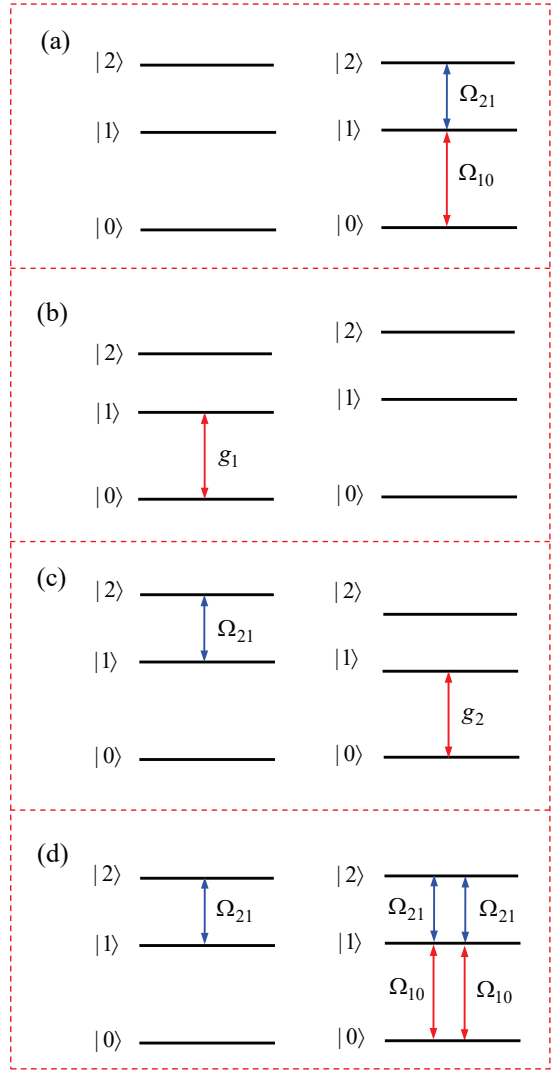} \vspace*{%
-0.08in}
\end{center}
\caption{(Color online) Illustration for the change of the level structure
of two transmon qutrits during the entanglement preparation. In (a), (b),
(c), and (d), figures on the left represent the level structure for qutrit 1
while figures on the right represent the level structure for qutrit 2. (a)
Decoupling of qutrits 1 and 2 from the cavity during applying microwave
pulses to qutrit 2. The pulse with a Rabi frequency $\mathrm{\Omega}_{21}$ is applied
after the pulse of a Rabi frequency $\mathrm{\Omega}_{10}$ is completed. (b)
Decoupling of qutrit 2 from the cavity during bringing the $\left\vert
0\right\rangle \leftrightarrow \left\vert 1\right\rangle $ transition of
qutrit 1 to resonance with the cavity. $g_1$ is the coupling constant. (c)
Decoupling of qutrit 1 from the cavity during applying a microwave pulse to
qutrit 1 and bringing the $\left\vert 0\right\rangle \leftrightarrow
\left\vert 1\right\rangle $ transition of qutrit 2 to resonance with the
cavity. $g_2$ is the coupling constant. (d) Decoupling of qutrits 1 and 2
from the cavity during applying microwave pulses to each qutrit (for the
sequence of the applied pulses, see the operation of step 5 in the text). In
(b) [(c)], during the cavity on resonance with the $\left\vert
0\right\rangle \leftrightarrow \left\vert 1\right\rangle $ transition of
qutrit 1 (2), the cavity is highly detuned (or decoupled) from the $%
\left\vert 1\right\rangle \leftrightarrow \left\vert 2\right\rangle $
transition of qutrit 1 (2) can be made by having the $\left\vert
1\right\rangle \leftrightarrow \left\vert 2\right\rangle $ transition
frequency highly detuning from the cavity frequency. Note that the coupling
or decoupling of the qutrits with the cavity can be achieved by adjusting
the level spacings of the qutrits [40-43].}
\label{fig:2}
\end{figure}

Let us now consider two transmon qutrits $1$ and $2$ embedded in a
single-mode cavity. Initially, the two qutrits are decoupled from the
cavity, qutrit $1$ is in the state $\alpha \left\vert 0\right\rangle +\gamma
\left\vert 1\right\rangle +\beta \left\vert 2\right\rangle ,$ qutrit $2$ is
in the ground state $\left\vert 0\right\rangle $, and the cavity is in the
vacuum state $\left\vert 0\right\rangle _{c}$. The initial state of qutrit $%
1 $ here can be prepared by applying classical microwave pulses. The initial
state of the whole system is thus given by $\left\vert \psi \right\rangle _{%
\mathrm{in}}=\left( \alpha \left\vert 0\right\rangle _{1}+\gamma \left\vert
1\right\rangle _{1}+\beta \left\vert 2\right\rangle _{1}\right) \left\vert
0\right\rangle _{2}\left\vert 0\right\rangle _{c}.$ Hereafter, the
subscripts 1 and 2 represent qutrits 1 and 2, respectively.

To begin with, we define $\omega _{20}$ ($\omega _{21}$) as the $\left\vert
0\right\rangle \leftrightarrow \left\vert 1\right\rangle $ ($\left\vert
1\right\rangle \leftrightarrow \left\vert 2\right\rangle $) transition
frequency of each qutrit and $\mathrm{\Omega }_{10}$ ( $\mathrm{\Omega }_{21}
$) as the pulse Rabi frequency of the coherent $\left\vert 0\right\rangle
\leftrightarrow \left\vert 1\right\rangle $ ($\left\vert 1\right\rangle
\leftrightarrow \left\vert 2\right\rangle $) transition. In addition, the
frequency, initial phase, and duration of a microwave pulse are denoted as \{%
$\omega ,\phi ,t^{\prime }$\} in the remaining of the paper.

The operations for creating the entangled state (3) are described below:

Step 1: Apply a microwave pulse of \{$\omega _{10},-\pi /2,\pi /\left( 4%
\mathrm{\Omega }_{10}\right) $\} and then a microwave pulse of \{$\omega
_{21},-\pi /2,\pi /\left( 2\mathrm{\Omega }_{21}\right) $\} to qutrit $2$
[Fig.~2(a)]. According to Eqs. (2) and (3), the pulse application results in
the state transformation $\left\vert 0\right\rangle _{2}\rightarrow $ $%
\left( \left\vert 0\right\rangle _{2}+\left\vert 2\right\rangle _{2}\right) /%
\sqrt{2}$ (see Appendix). Therefore, the initial state $\left\vert \psi
\right\rangle _{\mathrm{in}}$ of the system becomes
\begin{equation}
\frac{1}{\sqrt{2}}\left( \alpha \left\vert 0\right\rangle _{1}+\gamma
\left\vert 1\right\rangle _{1}+\beta \left\vert 2\right\rangle _{1}\right)
\left( \left\vert 0\right\rangle _{2}+\left\vert 2\right\rangle _{2}\right)
\left\vert 0\right\rangle _{c}.
\end{equation}

Step 2: Bring the $\left\vert 0\right\rangle \leftrightarrow \left\vert
1\right\rangle $ transition of qutrit 1 to resonance with the cavity for an
interaction time $\pi /\left( 2g_{1}\right) $ [Fig.~2(b)]. According to Eq.
(1), we have $\left\vert 1\right\rangle _{1}\left\vert 0\right\rangle
_{c}\rightarrow -i\left\vert 0\right\rangle _{1}\left\vert 1\right\rangle
_{c_{\backslash }}$. Thus, the state (4) changes to

\begin{equation}
\frac{1}{\sqrt{2}}\left[ \left( \alpha \left\vert 0\right\rangle _{1}+\beta
\left\vert 2\right\rangle _{1}\right) \left( \left\vert 0\right\rangle
_{2}+\left\vert 2\right\rangle _{2}\right) \left\vert 0\right\rangle
_{c}-i\gamma \left\vert 0\right\rangle _{1}\left( \left\vert 0\right\rangle
_{2}+\left\vert 2\right\rangle _{2}\right) \left\vert 1\right\rangle _{c}%
\right] .
\end{equation}%
\

Step 3: Bring the $\left\vert 0\right\rangle \leftrightarrow \left\vert
1\right\rangle $ transition of qutrit 2 to resonance with the cavity for an
interaction time $\pi /g_{2}$ [Fig.~2(c)]$,$ resulting in $\left\vert
0\right\rangle _{2}\left\vert 1\right\rangle _{c}\rightarrow -\left\vert
0\right\rangle _{2}\left\vert 1\right\rangle _{c}$ according to Eq. (1). The
state (5) thus changes to

\begin{equation}
\frac{1}{\sqrt{2}}\left[ \left( \alpha \left\vert 0\right\rangle _{1}+\beta
\left\vert 2\right\rangle _{1}\right) \left( \left\vert 0\right\rangle
_{2}+\left\vert 2\right\rangle _{2}\right) \left\vert 0\right\rangle
_{c}-i\gamma \left\vert 0\right\rangle _{1}\left( -\left\vert 0\right\rangle
_{2}+\left\vert 2\right\rangle _{2}\right) \left\vert 1\right\rangle _{c}%
\right] .
\end{equation}

Step 4: Bring the $\left\vert 0\right\rangle \leftrightarrow \left\vert
1\right\rangle $ transition of qutrit 1 to resonance with the cavity for an
interaction time $3\pi /\left( 2g_{1}\right) $ [Fig.~2(b)]. According to
Eq.~(1), we have $\left\vert 0\right\rangle _{1}\left\vert 1\right\rangle
_{c}\rightarrow i\left\vert 1\right\rangle _{1}\left\vert 0\right\rangle
_{c}.$ Hence, the state (6) changes to

\begin{equation}
\frac{1}{\sqrt{2}}\left[ \left( \alpha \left\vert 0\right\rangle _{1}+\beta
\left\vert 2\right\rangle _{1}\right) \left( \left\vert 0\right\rangle
_{2}+\left\vert 2\right\rangle _{2}\right) +\gamma \left\vert 1\right\rangle
_{1}\left( -\left\vert 0\right\rangle _{2}+\left\vert 2\right\rangle
_{2}\right) \right] \left\vert 0\right\rangle _{c}.
\end{equation}

Step 5: Apply microwave pulses of \{$\omega _{10},-\pi /2,\pi /\left( 2%
\mathrm{\Omega }_{10}\right) $\}, \{$\omega _{21},\pi /2,\pi /\left( 4%
\mathrm{\Omega }_{21}\right) $\}, \{$\omega _{10},\pi /2,3\pi /\left( 4%
\mathrm{\Omega }_{10}\right) $\}, and then \{$\omega _{21},\pi /2,\pi
/\left( 2\mathrm{\Omega }_{21}\right) $\} to qutrit 2 in turn [Fig.~2(d)].
According to Eqs. (2) and (3), the pulse application leads to $\left(
-\left\vert 0\right\rangle _{2}+\left\vert 2\right\rangle _{2}\right) /\sqrt{%
2}\rightarrow \left\vert 1\right\rangle _{2}$ but nothing to the state $%
\left( \left\vert 0\right\rangle _{2}+\left\vert 2\right\rangle _{2}\right) /%
\sqrt{2}$ of qutrit 2 (see Appendix). Meanwhile, apply a microwave pulse of
\{$\omega _{21},\pi /2,\pi /\left( 2\mathrm{\Omega }_{21}\right) $\} to
qutrit 1 [Fig.~2(d)], resulting in $\left\vert 2\right\rangle
_{1}\rightarrow \left\vert 1\right\rangle _{1}$ and $\left\vert
1\right\rangle _{1}\rightarrow -\left\vert 2\right\rangle _{1}$ according to
Eq. (3). Therefore, the state (7) becomes

\begin{equation}
\left[ \frac{1}{\sqrt{2}}\alpha \left( \left\vert 0\right\rangle _{1}+\beta
\left\vert 1\right\rangle _{1}\right) \left( \left\vert 0\right\rangle
_{2}+\left\vert 2\right\rangle _{2}\right) -\gamma \left\vert 2\right\rangle
_{1}\left\vert 1\right\rangle _{2}\right] \left\vert 0\right\rangle _{c}.
\end{equation}

Step 6: Bring the $\left\vert 0\right\rangle \leftrightarrow \left\vert
1\right\rangle $ transition of qutrit 1 to resonance with the cavity for an
interaction time $\pi /\left( 2g_{1}\right) $ [Fig.~2(b)]$.$ As a result, we
have $\left\vert 1\right\rangle _{1}\left\vert 0\right\rangle
_{c}\rightarrow -i\left\vert 0\right\rangle _{1}\left\vert 1\right\rangle
_{c}.$ Thus, the state (8) becomes

\begin{equation}
\frac{1}{\sqrt{2}}\left[ \alpha \left\vert 0\right\rangle _{1}\left(
\left\vert 0\right\rangle _{2}+\left\vert 2\right\rangle _{2}\right)
\left\vert 0\right\rangle _{c}-i\beta \left\vert 0\right\rangle _{1}\left(
\left\vert 0\right\rangle _{2}+\left\vert 2\right\rangle _{2}\right)
\left\vert 1\right\rangle _{c}\right] -\gamma \left\vert 2\right\rangle
_{1}\left\vert 1\right\rangle _{2}\left\vert 0\right\rangle _{c}.
\end{equation}

Step 7: Bring the $\left\vert 0\right\rangle \leftrightarrow \left\vert
1\right\rangle $ transition of qutrit 2 to resonance with the cavity for an
interaction time $\pi /g_{2}$ [Fig.~2(c)]$,$ resulting in $\left\vert
0\right\rangle _{2}\left\vert 1\right\rangle _{c}\rightarrow -\left\vert
0\right\rangle _{2}\left\vert 1\right\rangle _{c}$ and $\left\vert
1\right\rangle _{2}\left\vert 0\right\rangle _{c}\rightarrow -\left\vert
1\right\rangle _{2}\left\vert 0\right\rangle _{c}$. Meanwhile, apply a
microwave pulse of \{$\omega _{21},\pi /2,\pi /\mathrm{\Omega }_{21}$\} to
qutrit 1 [Fig.~2(c)], resulting in $\left\vert 2\right\rangle
_{1}\rightarrow $ $-\left\vert 2\right\rangle _{1}.$ Thus, the state (9)
becomes

\begin{equation}
\frac{1}{\sqrt{2}}\left[ \alpha \left\vert 0\right\rangle _{1}\left(
\left\vert 0\right\rangle _{2}+\left\vert 2\right\rangle _{2}\right)
\left\vert 0\right\rangle _{c}-i\beta \left\vert 0\right\rangle _{1}\left(
-\left\vert 0\right\rangle _{2}+\left\vert 2\right\rangle _{2}\right)
\left\vert 1\right\rangle _{c}\right] -\gamma \left\vert 2\right\rangle
_{1}\left\vert 1\right\rangle _{2}\left\vert 0\right\rangle _{c}.
\end{equation}

Step 8: Bring the $\left\vert 0\right\rangle \leftrightarrow \left\vert
1\right\rangle $ transition of qutrit 1 to resonance with the cavity for an
interaction time $3\pi /\left( 2g_{1}\right) $ [Fig.~2(b)]$.$ As a result,
we have $\left\vert 0\right\rangle _{1}\left\vert 1\right\rangle
_{c}\rightarrow i\left\vert 1\right\rangle _{1}\left\vert 0\right\rangle
_{c}.$Therefore, the state (10) changes to

\begin{equation}
\left\{ \frac{1}{\sqrt{2}}\left[ \alpha \left\vert 0\right\rangle _{1}\left(
\left\vert 0\right\rangle _{2}+\left\vert 2\right\rangle _{2}\right) +\beta
\left\vert 1\right\rangle _{1}\left( -\left\vert 0\right\rangle
_{2}+\left\vert 2\right\rangle _{2}\right) \right] -\gamma \left\vert
2\right\rangle _{1}\left\vert 1\right\rangle _{2}\right\} \left\vert
0\right\rangle _{c}.
\end{equation}

Step 9: Apply a microwave pulse of \{$\omega _{21},\pi /2,\pi /\left( 2%
\mathrm{\Omega }_{21}\right) $\} and then a microwave pulse of \{$\omega
_{10},\pi /2,\pi /\left( 4\mathrm{\Omega }_{10}\right) $\} to qutrit 2
[Fig.~2(a)], resulting in the state transformations $\left( \left\vert
0\right\rangle _{2}+\left\vert 2\right\rangle _{2}\right) /\sqrt{2}%
\rightarrow \left\vert 0\right\rangle _{2},$ $\left( -\left\vert
0\right\rangle _{2}+\left\vert 2\right\rangle _{2}\right) /\sqrt{2}%
\rightarrow \left\vert 1\right\rangle _{2},$ and $\left\vert 1\right\rangle
_{2}\rightarrow $ $-\left\vert 2\right\rangle _{2}$ (see Appendix). Hence,
the state (11) becomes

\begin{equation}
\left( \alpha \left\vert 0\right\rangle _{1}\left\vert 0\right\rangle
_{2}+\beta \left\vert 1\right\rangle _{1}\left\vert 1\right\rangle
_{2}+\gamma \left\vert 2\right\rangle _{1}\left\vert 2\right\rangle
_{2}\right) \left\vert 0\right\rangle _{c}.
\end{equation}

\begin{figure}[tbp]
\begin{center}
\includegraphics[bb=66 459 427 711, width=8.5 cm, clip]{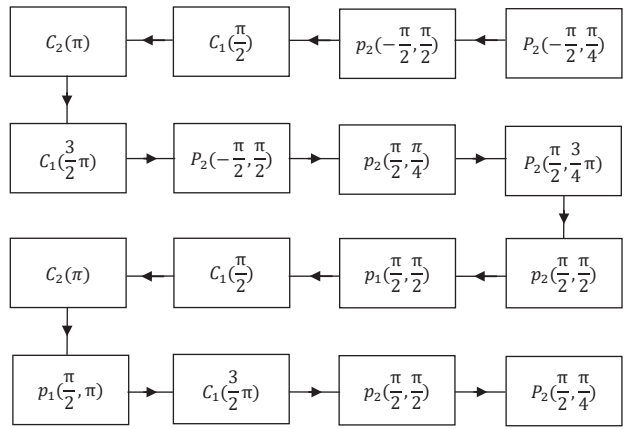} \vspace*{%
-0.08in}
\end{center}
\caption{(Color online) Circuit diagram for the two-qutrit entanglement
preparation (the operation sequence follows the arrows). The subscript $j$
of $C_{j},$ $p_{j},$ and $P_{j}$ represents qutrit $j$ ($j=1,2$) on which
the operation is performed. $C_{1}\left( g_{1}t\right) ,$ with $g_{1}t=%
\protect\pi /2$ or $3\protect\pi /2,$ represents the operation that the
cavity resonantly interacts with the $\left\vert 0\right\rangle
\leftrightarrow \left\vert 1\right\rangle $ transition of qutrit $1.$ $%
C_{2}\left( g_{2}t\right) ,$ with $g_{2}t=\protect\pi ,$ represents the
operation that the cavity resonantly interacts with the $\left\vert
0\right\rangle \leftrightarrow \left\vert 1\right\rangle $ transition of
qutrit $2.$ Here $g_{1}$ or $g_{2}$ is the qutrit-cavity coupling constant,
while $t$ is the qutrit-cavity interaction time. In addition, $p_{j}\left(
\protect\phi ,\mathrm{\Omega}_{21}t\right) ,$ with $\protect\phi =-\protect\pi /2$
or $\protect\pi /2$ and $\mathrm{\Omega}_{21}t=\protect\pi /4,$ $\protect\pi /4$ or $%
\protect\pi ,$ represents the operation that a microwave pulse is resonant
with the $\left\vert 1\right\rangle \leftrightarrow \left\vert
2\right\rangle $ transition of qutrit $j$ ($j=1,2$); while $P_{j}\left(
\protect\phi ,\mathrm{\Omega}_{10}t\right) ,$ with $\protect\phi =-\protect\pi /2$
or $\protect\pi /2$ and $\mathrm{\Omega}_{10}t=\protect\pi /4,$ $\protect\pi /2$ or $%
3\protect\pi /4,$ represents the operation that a microwave pulse is
resonant with the $\left\vert 0\right\rangle \leftrightarrow \left\vert
1\right\rangle $ transition of qutrit $j$ ($j=2$). Here, $\protect\phi $ is
the pulse initial phase, $\mathrm{\Omega}_{10}$ or $\mathrm{\Omega}_{21}$ is the pulse Rabi
frequency, while $t $ is the pulse duration.}
\label{fig:3}
\end{figure}

The above operations are illustrated in Fig. 3. The result (12) shows that
the two qutrits are prepared in an entangled state%
\begin{equation}
\left\vert \psi \right\rangle _{\mathrm{ent}}=\alpha \left\vert
0\right\rangle _{1}\left\vert 0\right\rangle _{2}+\beta \left\vert
1\right\rangle _{1}\left\vert 1\right\rangle _{2}+\gamma \left\vert
2\right\rangle _{1}\left\vert 2\right\rangle _{2},
\end{equation}%
while the cavity returns to its original vacuum state after the entire
operation. It can be seen that the prepared entangled state (13) is
different from the two-qubit entangled states $\alpha \left\vert
00\right\rangle +\beta \left\vert 11\right\rangle /\sqrt{2}$ and $\alpha
\left\vert 01\right\rangle +\beta \left\vert 10\right\rangle .$ It is well
known that the two-qubit entangled states have been experimentally generated
with various physical systems. Interestingly, it is noted that the procedure
for entangling the two qutrits has nothing to do with $\alpha ,\beta $, and $%
\gamma $. Moreover, the degree of entanglement for the prepared two-qutrit
entangled state (13) is controllable by varying the weight factors $\alpha ,$
$\beta ,$ and $\gamma $ involved in the initial state of qutrit 1. For $%
\alpha =\beta =\gamma =1/\sqrt{3},$ the prepared entangled state (13) is a
maximally entangled state $\left( \left\vert 0\right\rangle _{1}\left\vert
0\right\rangle _{2}+\left\vert 1\right\rangle _{1}\left\vert 1\right\rangle
_{2}+\left\vert 2\right\rangle _{1}\left\vert 2\right\rangle _{2}\right) /%
\sqrt{3}$ of the two qutrits. Otherwise, the two qutrits are prepared in a
partially entangled state.

From description given above, one can see that this proposal does not
require that the qutrit-cavity coupling constants $g_{1}$ and $g_{2}$ are
identical. Thus, either non-uniformity in the qutrit device parameters or
non-exact placement of qutrits in the cavity is allowed, which significantly
reduces the experiemental engineering difficulty. In addition, neither
auxiliary qutrit nor measurement is needed for the two-qutrit entanglement
production. Furthermore, as shown above, this proposal only requires two
different microwave-pulse frequencies ($\omega _{10}$ and $\omega _{21}$),
which are readily achievable in current experiments [41].

Before ending this section, several points are addressed as follows:

(a) During the operations of steps (2-4) and (6-8), the $\left\vert
0\right\rangle \leftrightarrow \left\vert 1\right\rangle $ transition of
qutrit 1 or 2 is on resonance with the cavity. This can be achieved by
adjusting the level spacings of the qutrits. For superconducting qutrits,
the level spacings can be rapidly (within a few nanoseconds) adjusted by
varying external control parameters (e.g., the magnetic flux applied to a
superconducting loop of phase, transmon, Xmon, or flux qubits; see, e.g.,
[42-45]).

(b). Durng the cavity interacting with a qutrit, the other qutrit is
decoupled from the cavity. During the pulse application, both qutrits are
decoupled from the cavity, in order to avoid the effect of the unwanted
qutrit-cavity interactions. In addition, during the cavity on resonance with
the $\left\vert 0\right\rangle \leftrightarrow \left\vert 1\right\rangle $
transition of a qutrit, the $\left\vert 1\right\rangle \leftrightarrow $ $%
\left\vert 2\right\rangle $ transition of this qutrit is highly detuned (or
decoupled) from the cavity. In principle, these conditions can be satisfied
by adjusting the level spacings of the qutrits [42-45].

(c). The ladder-type level structure is available in natural atoms or other
artificial atoms such as nitrogen-vacancy (NV) centers, quantum dots, and
various other SC qutrits (SC phase, charge, and Xmon qubits). For NV
centers, the level spacings can be readily adjusted by changing the external
magnetic field applied along the crystalline axis of each NV center [46,47].
For quantum dots and atoms, the level spacings can be adjusted by changing
the voltage on the electrodes around each atom/quantum dot [48]. Thus, this
proposal can be applied to create the two-qutrit entangled state (13) in a
wide range of physical systems.

(d). The method presented here is applicable to a 1D, 2D, or 3D cavity or
resonator as long as the conditions described above can be met.

\begin{center}
\textbf{IV. DISCUSSION AND POSSIBLE EXPERIMENTAL IMPLEMENTATION}
\end{center}

The total operation time is
\begin{equation}
\tau =\frac{7\pi }{4\mathrm{\Omega }_{10}}+\frac{13\pi }{4\mathrm{\Omega }%
_{21}}+\frac{4\pi }{g_{1}}+\frac{2\pi }{g_{2}}+8\tau _{d},
\end{equation}%
where $t_{d}$ is the typical time required for adjusting the qutrit level
spacings. To reduce the effect of decoherence from the qutrits, the
operation time $\tau $ should be much smaller than the energy relaxation
time and the dephasing time of qutrits. In addition, to reduce the effect of
dissipation from the cavity, $\tau $ should be much smaller than the
lifetime of the cavity mode, which is given by $\kappa ^{-1}=Q/\left( 2\pi
\nu _{c}\right) $ ($Q$ is the quality factor of the cavity while $\nu _{c}$
is the cavity frequency). In principle, these requirements can be satisfied
for the following reasons. First, the operation time $\tau $ can be
shortened by increasing the coupling constants $g_{1}$ and $g_{2}$ (e.g.,
placing each qutrit at an antinode of the cavity magnetic field), increasing
the pulse Rabi frequencies $\mathrm{\Omega }_{10}$ and $\mathrm{\Omega }_{21}
$ (e.g., through increasing the pulse intensity), and rapidly adjusting the
qutrit level spacings (e.g., $\tau _{d}$ $\sim $ $1-3$ ns is the typical
time for adjusting the level spacings of a superconducting qutrit in
experiments [41,43,49]). Second, one can choose qutrits with long
decoherence times. Last, $\kappa ^{-1}$ can be increased by employing a high-%
$Q$ resonator.

Let us now consider the experimental possibility of preparing the two-qutrit
entangled state (13) with two superconducting transmon qutrits embedded in a
one-dimensional transmission line resonator (TLR) (Fig. 4).

\begin{figure}[tbp]
\begin{center}
\includegraphics[bb=105 578 492 739, width=8.5 cm, clip]{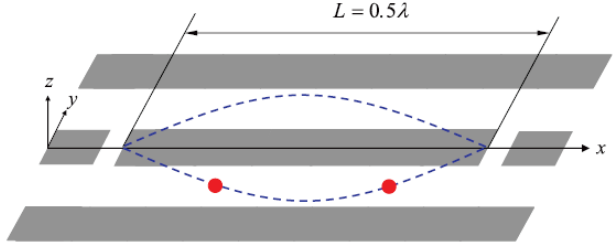} \vspace*{%
-0.08in}
\end{center}
\caption{(Color online) Setup for two transmon qutrits (red dots) embedded
in a one-dimensional coplanar waveguide resonator. $\protect\lambda$ is the
wavelength of the resonator mode, and $L$ is the length of the resonator.
The two (blue) curved lines represent the standing wave magnetic field of
the resonator in the $z$ direction, which is perpendicular to the surface of
the superconducting loop of each transmon qutrit. The two transmon qutrits
are placed at locations where the magnetic fields are the same to achieve an
identical coupling strength for each qutrit. The superconducting loop (not
shown) of each transmon qutrit is located in the plane of the resonator
between the two lateral ground planes (i.e., the x-y plane).}
\end{figure}

When dissipation and dephasing are considered, the dynamics of the lossy
system is determined by
\begin{align}
\frac{d\rho }{dt}=& -i[H_{\mathrm{I}},\rho ]+\kappa \mathcal{L}[a]  \notag \\
& +\sum\limits_{j=1,2}\gamma _{10,j}\mathcal{L}[\sigma _{10,j}^{-}]+\gamma
_{21,j}\mathcal{L}[\sigma _{21,j}^{-}]+\gamma _{20,j}\mathcal{L}[\sigma
_{20,j}^{-}]  \notag \\
& +\sum\limits_{j=1,2}\{\gamma _{\varphi 1,j}(\sigma _{11,j}\rho \sigma
_{11,j}-\sigma _{11,j}\rho /2-\rho \sigma _{11,j}/2)\}  \notag \\
& +\sum\limits_{j=1,2}\{\gamma _{\varphi 2,j}(\sigma _{22,j}\rho \sigma
_{22,j}-\sigma _{22,j}\rho /2-\rho \sigma _{22,j}/2)\},
\end{align}%
where $H_{\mathrm{I}}$ is the Hamiltonian $H_{\mathrm{I}_{1}},H_{\mathrm{I}%
_{2}},$or $H_{\mathrm{I}_{3}}$ above, the subscript $j$ represents qutrit $j$
($j=1,2$); $\sigma _{10,j}^{-}=|0\rangle _{j}\langle 1|$, $\sigma
_{21,j}^{-}=|1\rangle _{j}\langle 2|$, $\sigma _{20,j}^{-}=|0\rangle
_{j}\langle 2|$, $\sigma _{11,j}=|1\rangle _{j}\langle 1|,\sigma
_{22,j}=|2\rangle _{j}\langle 2|$; and $\mathcal{L}[\xi ]=\xi \rho \xi
^{\dag }-\xi ^{\dag }\xi \rho /2-\rho \xi ^{\dag }\xi /2$, with $\xi
=a,\sigma _{10,j}^{-},\sigma _{21,j}^{-},\sigma _{20,j}^{-}$. Here, $\kappa $
is the photon decay rate of the cavity or resonator. In addition, $\gamma
_{10,j}$ is the energy relaxation rate for the level $|1\rangle $ of qutrit $%
j,$ $\gamma _{21,j}$ $(\gamma _{20,j})$ is the energy relaxation rate of the
level $|2\rangle $ of qutrit $j$ for the decay path $|2\rangle
\longrightarrow |1\rangle (|0\rangle )$, and $\gamma _{\varphi 1,j}$ $\left(
\gamma _{\varphi 2,j}\right) $ is the dephasing rate of the level $|1\rangle
$ $(|2\rangle )$ of qutrit $j$ ($j=1,2$). \newline

The fidelity of the operations is given by
\begin{equation}
\mathcal{F}=\sqrt{\langle \psi _{\mathrm{id}}|\rho |\psi _{\mathrm{id}%
}\rangle },
\end{equation}%
where $|\psi _{\mathrm{id}}\rangle $ is the output state of an ideal system
without dissipation and dephasing; while $\rho $ is the final practical
density operator of the system when the operation is performed in a
realistic situation. As an example, consider $\alpha =\beta =\gamma =1/\sqrt{%
3},$ for which the prepared entangled state (13) is $|\psi _{\mathrm{id}%
}\rangle =\left( \left\vert 0\right\rangle _{1}\left\vert 0\right\rangle
_{2}+\left\vert 1\right\rangle _{1}\left\vert 1\right\rangle _{2}+\left\vert
2\right\rangle _{1}\left\vert 2\right\rangle _{2}\right) /\sqrt{3}$ for the
ideal case, which corresponds to the case that the initial state of the
whole system is $\left\vert \psi \right\rangle _{\mathrm{in}}=1/\sqrt{3}%
\left( \alpha \left\vert 0\right\rangle _{1}+\gamma \left\vert
1\right\rangle _{1}+\beta \left\vert 2\right\rangle _{1}\right) \left\vert
0\right\rangle _{2}\left\vert 0\right\rangle _{c}$.

\begin{figure}[tbp]
\begin{center}
\includegraphics[bb=32 12 601 316, width=8.5 cm, clip]{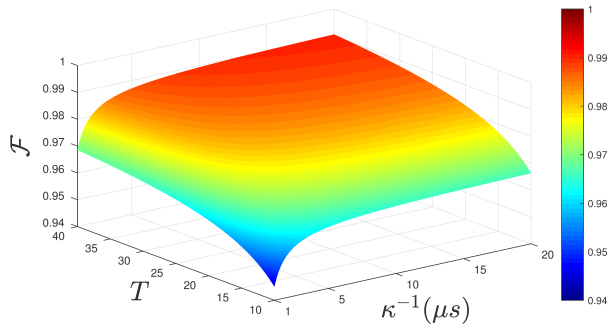} \vspace*{%
-0.08in}
\end{center}
\caption{(Color online) Fidelity versus $T$ and $\protect\kappa^{-1}$. The
parameters used in the numerical simulation are referred to the text.}
\label{fig:5}
\end{figure}

The parameters used in the numerical simulation are: (i) $\gamma
_{10,j}^{-1}=2T$ $\mu $s, $\gamma _{20,j}^{-1}=5T$ $\mu $s, $\gamma
_{21,j}^{-1}=T$ $\mu $s, $\gamma _{\phi 1,j}^{-1}=\gamma _{\phi 2,j}^{-1}=T$
$\mu $s, (ii) $g_{1},$ $g_{2},$ $\mathrm{\Omega }_{10},$ $\mathrm{\Omega }%
_{21}=2\pi \times 100$ MHz. The coupling strengths and the pulse Rabi
frequencies with the values chosen here are readily available in experiments
because a coupling strength $g/2\pi \sim 220$ MHz has been experimentally
demonstrated for a transmon device coupled to a TLR [50] and a microwave
pulse Rabi frequency $\mathrm{\Omega }/2\pi \sim 300$ MHz has been reported
in experiments [51].

By solving the master equation (15), we numerically calculate the fidelity
versus $T$ and $\kappa ^{-1}$, as shown in Fig.~5. Figure 5 illustrates that
when $T=30$ $\mu $s and $\kappa ^{-1}\geq 2.5$ $\mu $s, fidelity exceeds $%
98.03\%.$ This result implies that a high fidelity can be obtained when the
dissipation and dephasing are taken into account.

For $T=30$ $\mu $s, we have $\gamma _{10,j}^{-1}=60$ $\mu $s, $\gamma
_{20,j}^{-1}=150$ $\mu $s, $\gamma _{21,j}^{-1}=30$ $\mu $s, and $\gamma
_{\phi 1,j}^{-1}=\gamma _{\phi 2,j}^{-1}=30$ $\mu $s, which are much longer
than the total operation time $\sim 67$ ns (estimated based on the coupling
constants and the pulse Rabi frequencies chosen above and for $\tau _{d}\sim
1.5$ ns). Note that energy relaxation time and dephasing time can be made to
be on the order of $25-100$ $\mu $s for state-of-the-art superconducting
transmon devices at the present time [52-55]. It is worth noting that for an
ideal transmon, the $\left\vert 0\right\rangle $ $\leftrightarrow \left\vert
2\right\rangle $ transition is theoretically zero due to the selection rule
[40,56]; and in practice, there exists a weak transition between the two
levels $\left\vert 0\right\rangle $ and $\left\vert 2\right\rangle $ [57].
Thus, $\gamma _{20,j}^{-1}\gg \gamma _{10,j}^{-1},\gamma _{21,j}^{-1}.$

For a transmon qutrit, the typical transition frequency between the two
lowest levels $\left\vert 0\right\rangle $ and $\left\vert 1\right\rangle $
can be made to be $3-10$ GHz. Thus, as an example, choose $\nu _{c}\sim 3$
GHz. For $\kappa ^{-1}=2.5$ $\mu $s, we have $Q\sim $ $4.7\times 10^{4}.$
The required resonator quality factor here is achievable in experiments
because TLRs with a (loaded)\ quality factor $Q\sim 10^{6}$ have been
experimentally demonstrated [58,59].

The analysis presented above shows that high-fidelity generation of the
two-qutrit entangled state is feasible within present-day circuit QED.

Finally, for two superconducting qutrits located in a microwave resonator,
they can be well separated, because the dimension of a superconducting
qutrit is 10 to 100 micrometers while the wavelength of the cavity mode for
a microwave superconducting resonator is 1 to a few centimeters [60,61]. As
long as the two qutrits are well separated in space (Fig.~4), the loop
current of one qutrit affecting the other qutrit and the direct coupling
between the two qutrits are negligible, which can be reached by designing
the qutrits and the resonator appropriately.

\begin{center}
\textbf{V. CONCLUSION}
\end{center}

We have presented a method to entangle two transmon qutrits based on circuit
QED. Since only resonant interactions are employed, the entanglement can be
produced within a short time (a few tens of nanoseconds) and thus
decoherence from the qutrits and the cavity are greatly suppressed. As shown
above, this proposal does not require measurement on the states of the
qutrits or the cavity-mode photons and only requires resonant qutrit-cavity
interaction and resonant qutrit-pulse interaction for each step of the
operation. Thus, this proposal is easy to implement in experiments. This
proposal is quite general and can be applied to accomplish the same task
with a wide range of physical systems. To the best of our knowledge, our
work is the first to show how to create entangled states of qutrits with
natural or artificial atoms based on cavity or circuit QED. We remark that
the number of pulses can be reduced by employing dispersive qutrit-cavity
interaction but the operation time will become much prolonged due to the
large detuning technique to be used. We hope that this work will stimulate
the experimental activities in the near future.

\begin{center}
\textbf{ACKNOWLEDGMENTS}
\end{center}

This work was supported in part by the Natural Science Foundation of Fujian
Province of China under Grant No. 2015J01020, the Zhejiang Natural Science
Foundation under Grant No. LZ13A040002, the NKRDP of China (Grant No.
2016YFA0301802), and the National Natural Science Foundation of China under
Grant Nos. [11074062, 11374083, 11774076].

\begin{center}
\textbf{APPENDIX}
\end{center}

We here give a derivation on the state transformations induced by the
application of the pulses, for the operations of steps 1, 5, and 9 described
above.

For step 1: A microwave pulse of \{$\omega _{10},-\pi /2,\pi /\left( 4%
\mathrm{\Omega }_{10}\right) $\} and then a microwave pulse of \{$\omega
_{21},-\pi /2,\pi /\left( 2\mathrm{\Omega }_{21}\right) $\} were applied to
qutrit $2$ [Fig.~2(a)]. The first pulse results in $\left\vert
0\right\rangle _{2}\rightarrow \left( \left\vert 0\right\rangle
_{2}+\left\vert 1\right\rangle _{2}\right) /\sqrt{2}$ according to Eq. (2),
while the second pulse leads to $\left\vert 1\right\rangle _{2}\rightarrow
\left\vert 2\right\rangle _{2}$ according to Eq. (3). After the two pulses,
we thus have the state transformations:%
\begin{equation}
\left\vert 0\right\rangle _{2}\overset{p1}{\rightarrow }\left( \left\vert
0\right\rangle _{2}+\left\vert 1\right\rangle _{2}\right) /\sqrt{2}\overset{%
p2}{\rightarrow }\left( \left\vert 0\right\rangle _{2}+\left\vert
2\right\rangle _{2}\right) /\sqrt{2},
\end{equation}%
where the first transformation is obtained after applying the first pulse
while the second transformation is achieved after applying the second pulse.
Note that the state $\left\vert 0\right\rangle _{2}$ remains unchanged
during the second pulse because the pulse is highly detuned (or decoupled)
from the $\left\vert 1\right\rangle _{2}\rightarrow \left\vert
2\right\rangle _{2}$ transition of qutrit 2.

For step 5: The microwave pulses of \{$\omega _{10},-\pi /2,\pi /\left( 2%
\mathrm{\Omega }_{10}\right) $\}, \{$\omega _{21},\pi /2,\pi /\left( 4%
\mathrm{\Omega }_{21}\right) $\}, \{$\omega _{10},\pi /2,3\pi /\left( 4%
\mathrm{\Omega }_{10}\right) $\}, and then \{$\omega _{21},\pi /2,\pi
/\left( 2\mathrm{\Omega }_{21}\right) $\} were applied to qutrit 2 in turn
[Fig.~2(d)]. According to Eqs. (2) and (3), the first pulse results in $%
\left\vert 0\right\rangle _{2}\rightarrow \left\vert 1\right\rangle _{2},$
the second pulse results in $\left( \left\vert 1\right\rangle
_{2}+\left\vert 2\right\rangle _{2}\right) /\sqrt{2}\rightarrow \left\vert
1\right\rangle _{2}$ and $\left( -\left\vert 1\right\rangle _{2}+\left\vert
2\right\rangle _{2}\right) /\sqrt{2}\rightarrow \left\vert 2\right\rangle
_{2},$ the third pulse results in $\left\vert 1\right\rangle _{2}\rightarrow
\left( \left\vert 0\right\rangle _{2}-\left\vert 1\right\rangle _{2}\right) /%
\sqrt{2}$ but nothing to the state $\left\vert 2\right\rangle _{2}$, and the
last pulse leads to $\left\vert 1\right\rangle _{2}\rightarrow -\left\vert
2\right\rangle _{2}$ and $\left\vert 2\right\rangle _{2}\rightarrow
\left\vert 1\right\rangle _{2}.$ Based on these results, we can obtain the
following state transformations:

\begin{eqnarray}
&&\left( \left\vert 0\right\rangle _{2}+\left\vert 2\right\rangle
_{2}\right) /\sqrt{2}\overset{p1}{\rightarrow }\left( \left\vert
1\right\rangle _{2}+\left\vert 2\right\rangle _{2}\right) /\sqrt{2}\overset{%
p2}{\rightarrow }\left\vert 1\right\rangle _{2}\overset{p3}{\rightarrow }%
\left( \left\vert 0\right\rangle _{2}-\left\vert 1\right\rangle _{2}\right) /%
\sqrt{2}\overset{p4}{\rightarrow }\left( \left\vert 0\right\rangle
_{2}+\left\vert 2\right\rangle _{2}\right) /\sqrt{2},  \notag \\
&&\left( -\left\vert 0\right\rangle _{2}+\left\vert 2\right\rangle
_{2}\right) /\sqrt{2}\overset{p1}{\rightarrow }\left( -\left\vert
1\right\rangle _{2}+\left\vert 2\right\rangle _{2}\right) /\sqrt{2}\overset{%
p2}{\rightarrow }\left\vert 2\right\rangle _{2}\overset{p3}{\rightarrow }%
\left\vert 2\right\rangle _{2}\overset{p4}{\rightarrow }\left\vert
1\right\rangle _{2},
\end{eqnarray}%
where the first, second, third, and last transformations are obtained after
applying the first, second, third, and the last pulses, respectively.

\bigskip For step 9: A microwave pulse of \{$\omega _{21},\pi /2,\pi /\left(
2\mathrm{\Omega }_{21}\right) $\} and then a microwave pulse of \{$\omega
_{10},\pi /2,\pi /\left( 4\mathrm{\Omega }_{10}\right) $\} were applied to
qutrit 2 [Fig.~2(a)]. The first pulse results in $\left\vert 2\right\rangle
_{2}\rightarrow $ $\left\vert 1\right\rangle _{2}$ and $\left\vert
1\right\rangle _{2}\rightarrow $ $-\left\vert 2\right\rangle _{2}$ while the
second pulse leads to $\left( \left\vert 0\right\rangle _{2}+\left\vert
1\right\rangle _{2}\right) /\sqrt{2}\rightarrow $ $\left\vert 0\right\rangle
_{2}$ and $\left( -\left\vert 0\right\rangle _{2}+\left\vert 1\right\rangle
_{2}\right) /\sqrt{2}\rightarrow \left\vert 1\right\rangle _{2}$ but nothing
to the state $\left\vert 2\right\rangle _{2}.$ Based on these results, one
can have the following state transformations:%
\begin{eqnarray}
&&\left( \left\vert 0\right\rangle _{2}+\left\vert 2\right\rangle
_{2}\right) /\sqrt{2}\overset{p1}{\rightarrow }\left( \left\vert
0\right\rangle _{2}+\left\vert 1\right\rangle _{2}\right) /\sqrt{2}\overset{%
p2}{\rightarrow }\left\vert 0\right\rangle _{2},  \notag \\
&&\left( -\left\vert 0\right\rangle _{2}+\left\vert 2\right\rangle
_{2}\right) /\sqrt{2}\overset{p1}{\rightarrow }\left( -\left\vert
0\right\rangle _{2}+\left\vert 1\right\rangle _{2}\right) /\sqrt{2}\overset{%
p2}{\rightarrow }\left\vert 1\right\rangle _{2},  \notag \\
&&\left\vert 1\right\rangle _{2}\overset{p1}{\rightarrow }-\left\vert
2\right\rangle _{2}\overset{p2}{\rightarrow }-\left\vert 2\right\rangle _{2},
\end{eqnarray}%
where the first transformation is obtained after the first pulse while the
second transformation is achieved after the second pulse.

Note that the $p1,$ $p2,$ $p3,$ and $p4$ above represent the first, second,
third, and fourth pulses, respectively.

\section{\protect\bigskip Confict of Interest:}

The authors declare that they have no confict of interest.

\end{document}